\documentstyle[wor-sci,epsf]{article}
\begin{titlepage}
   \title{Relativistic Charge Form Factor of the Deuteron}
     \author{Andrei V.~Afanasev  \\ {\it North Carolina Central University, 
   Durham, NC 27707} \\ {\it and} \\ {\it Thomas Jefferson National Accelerator Facility,
   Newport News, VA 23606}
  \and
 V.D.~Afanas'ev, S.V.~Trubnikov \\
{\it Kharkov State University, 310077 Kharkov, Ukraine} }
 \end{titlepage}
\begin{document}
\maketitle

\begin{abstract}
      Relativistic integral representation in terms of experimental
neutron--proton scattering phase shifts  alone is used to compute 
the charge form factor of the deuteron $ G_{Cd}(Q^2)$. The results of numerical
 calculations of
$ |G_{Cd}(Q^2)|$ are presented in the interval of the
four--momentum transfers squared \mbox{$ 0 \leq Q^2 \leq 35\:fm^{-2}$.}
Zero and the prominent secondary maximum in  $ |G_{Cd}(Q^2)|$ are the
       direct consequences of the change of sign in the experimental 
$^3S_1 $-- phase shifts. Till the point $Q^2 \simeq 20 \;fm^{-2}$  the
total relativistic correction to $ |G_{Cd}(Q^2)|$ is positive and reaches
the maximal value of 25\% at  $ Q^2 \simeq 14 fm^{-2}$.
 \end{abstract}
\newpage
\begin{sloppypar}
      Deuteron is the  brightest example of intersection of nuclear and
particle physics. During more then sixty years it serves as source of
important information about the nuclear forces, 
mesonic and  baryonic degrees of freedoms in nuclei, relativistic effects 
 and a possible role of  quarks  
in nuclear structure. Therefore it is not surprising that currently
the electromagnetic (EM) structure of the deuteron is a subject of
 intensive theoretical (the list of publication is immense) and
experimental investigations.

With new experimental data from Jefferson Lab on elastic electron-deuteron scattering
 expected in the near future \cite{betsy, makis}, at momentum transfers in the GeV-range,
 one needs to develop relativistic approaches to the (np)-bound state problem.
 Recent experimental results from MIT-Bates \cite{bates} provided the first experimental
evidence for a zero in the deuteron charge form factor $G_{Cd}$  at about $Q^2$= 20 fm$^{-2}$ 
predicted in a number of theoretical models (or not predicted, as in some kinds of quark models).
 Here we report  new results of numerical calculations of   $G_{Cd}$ . These calculations are
based on the approach to the relativistic impulse  approximation, which was
briefly discussed in ref.~\cite{two} (see also the review~\cite{three}
and, especially, the references herein).  The more detailed formulae are
contained in ref.~\cite{four}.  In this approach the deuteron form factors are expressed
in terms of experimental neutron--proton $(n-p)$ phase shifts in the
triplet scattering channel   and experimental values of nucleon
EM form factors.

        According to ref. \cite{four} the formula for $G_{Cd}(Q^2)$
appears  as
 \begin{eqnarray}
G_{Cd}(Q^2)& = & (\rho \tilde{B}^{20} +\tilde{B}^{22})^2 G_{Cd}^{00}(Q^2) -
\nonumber\\
&&-(\rho \tilde{B}^{20} + \tilde{B}^{22})(\rho \tilde{B}^{00} +
\tilde{B}^{02}) [G_{Cd}^{02}(Q^2) + G_{Cd}^{20}(Q^2) ] + \nonumber\\
&&+ (\rho \tilde{B}^{00} + \tilde{B}^{02})^2 G_{Cd}^{22}(Q^2)\;.
\label{one}
\end{eqnarray}

       In  eq.(~\ref{one})  $\rho  $    is  the  constant  which  describes
  mixing of two $n-p$ states  with different orbital moments ($l=0$   and
 $l=2$) at  the point  of  the bound  state, {\it i.e.},  the  deuteron. This constant
 is  defined   by    the    correspondence    principle.    Analysing    the
nonrelativistic limits of  eqs.(1),(2), we can  prove that $\rho   $ appears
to be the standard  asymptotic $D/S$ --ratio  of the  radial deuteron   wave
functions,   so $   \rho =   0.0277$ (numerical  calculations show  that the
 dependence of DCFF on the variation  of $\rho $  is very  weak). All   four
elements  of the  matrix $ \tilde{B}^{ll'}(s)$ ($ l,l' = 0,2$)~\footnote{
 For the  choice of kinematic   variables here and in eq.(2)  see Appendix A.}
are taken  at  the  bound  state  point  $  s=M_d^2$  ($M_d =  2M
-\varepsilon  $, where  $M_d,M$    are  deuteron and nucleon masses and
  $\varepsilon $ is the  deuteron  binding    energy).  All  relativistic
 aspects   of   the two--nucleon problem are contained in $G^{ll'}$--
 matrix:

\begin{eqnarray}
 G_{Cd}^{ll'}&=&\Gamma ^2 \int_{4M^2}^{\infty} \frac{ds\Delta B^\dagger(s)}
 {s - M^2_d}\int_{s_1(s,t)}^{s_2(s,t)} \frac{ds' g_c(s,s't) \Delta B(s')}
{s'-M^2_d}, \nonumber\\
s_{2,1}&=&2M^2 + \frac{1}{2M^2}(2M^2 -t)\cdot(s-2M^2) \pm \nonumber\\
&&\pm \frac{1}{2M^2} \sqrt{(-t)(4M^2-t)s(s-4M^2)}\;.
\label{two}
\end{eqnarray}

        In eq.(\ref{two}) $\Gamma ^2$ is the normalization constant,
 which is calculated from the condition  $G_{Cd}(0) =1$. Matrix
 functions $\Delta B^{ll'}(s) = B^{ll'}(s + i \varepsilon ) - B^{ll'}
 (s-i\varepsilon )$ are the discontinuities of the Jost matrix
 $B(s)$. As usual, the Jost matrix  is the solution of the
 boundary problem in two--channel scattering theory:

\begin{eqnarray}
 S(s)B_{+}(s)& =& B_{-}(s), \nonumber\\
 s &\geq& 4M^2,
\label{three}
\end{eqnarray}
\begin{equation}
  S(s) \equiv S \left[ \delta ,\eta ,\varepsilon \right]= \left(
  \begin{array}{ll}
  \cos{2\varepsilon }\cdot e^{2i \delta }& i\sin {2\varepsilon } \cdot
  e^{i( \delta + \eta )}\\
  i\sin {2\varepsilon } \cdot e^ {i( \delta + \eta )} &
  \cos{2\varepsilon }\cdot e^ {2i \eta }
  \end{array}
\right)\;.
\label{four}
 \end{equation}

   The reduced Jost matrix $\tilde{B}$ in
      eq.(~\ref{one}) is the solution of the same eq.(\ref{three}) with the
scattering matrix $ \tilde{S}\equiv S(\tilde{\delta }, \tilde{\varepsilon
},\tilde{ \eta })$. Expressions  for $\tilde{B}$ and $B$ in terms of
  $n-p$ phase shifts are cumbersome and are summarized in Appendix B.

\begin{figure}[t]
\let\picnaturalsize=N
\def\picsize{3in}
\def\picfilenamea{fig1.eps}
\ifx\nopictures Y\else{\ifx\epsfloaded Y\else\input epsf \fi
\let\epsfloaded=Y
\centerline{
\ifx\picnaturalsize N\epsfxsize \picsize\fi \epsfbox{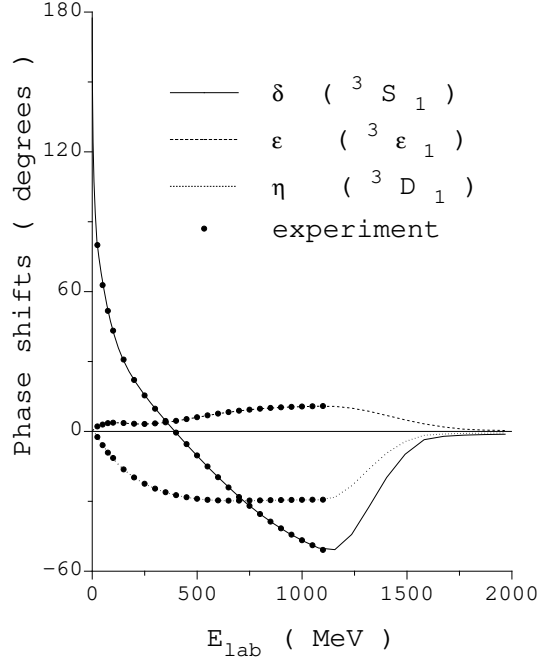}}}
\fi
\caption{ Neutron--proton phase shifts $^3S_1, ^3D_1, ^3 \varepsilon_1$ used in the calculations.
     Experimental data are taken from the VPI  analysis ref.\cite{Virginia}.}
\label{fig1}
\end{figure}

        The matrix functions $ g^{ll'}_c(s,s',t)$ of three variables are
  the relativistic charge form factors of the unconnected part of the matrix element of EM
current $\langle n'p'|j_\mu |np\rangle$. The results of the calculations of
  $ g^{ll'}_c$ are given in Appendix C. It is interesting to note that in the
  general case in the relativistic regime $ g^{ll'}_c$-- functions are not
  factorizable in $ s,s'$ variables, whereas in the nonrelativistic limit such
  factorization takes place. It means that in the framework of the  used
  relativistic approach \cite{two}-\cite{four} it is impossible to introduce a concept
   of relativistic deuteron wave function.

  The experimental set of $n-p$ phase shifts were taken from the analysis of
Virginia Tech group~\cite{Virginia} and is shown in Fig.~\ref{fig1}. This analysis
was made in the energy range \mbox{$ 0 < E_{lab} \leq 1100 $ MeV}.
  Extrapolation to higher energies is not as  important for the
  calculations of $G_{Cd}$  for the small and intermediate values of $Q^2$.
  The only essential circumstance is  that $^3S_1$--phase shifts change  sign from
  positive to negative and have the minimum near the energy $E_{lab} \sim 1
  $GeV, then go to zero in accordance with the Levinson's theorem. Any
  realistic $n-p  $ $ ^3S_1$ -- phase shift analysis has such a
  behavior.  Two other states ($^3D_1$ and $^3 \varepsilon _1$ ) give a
  relatively small contribution to $G_{Cd}$.

      For the calculations of $G_{Cd}$ we used (as a first step) the simplest
  choice of the nucleon form factors: $ G_{Ep} = (1+Q^2/18.23\; fm^{-2})^{-2},
  G_{Mp}/\mu _p = G_{Mn}/\mu _n = G_{Ep},\; G_{En} \equiv 0 $ for all
  $ Q^2$.

\begin{figure}[t]
\let\picnaturalsize=N
\def\picsize{3in}
\def\picfilenamea{fig2.eps}
\ifx\nopictures Y\else{\ifx\epsfloaded Y\else\input epsf \fi
\let\epsfloaded=Y
\centerline{
\ifx\picnaturalsize N\epsfxsize \picsize\fi \epsfbox{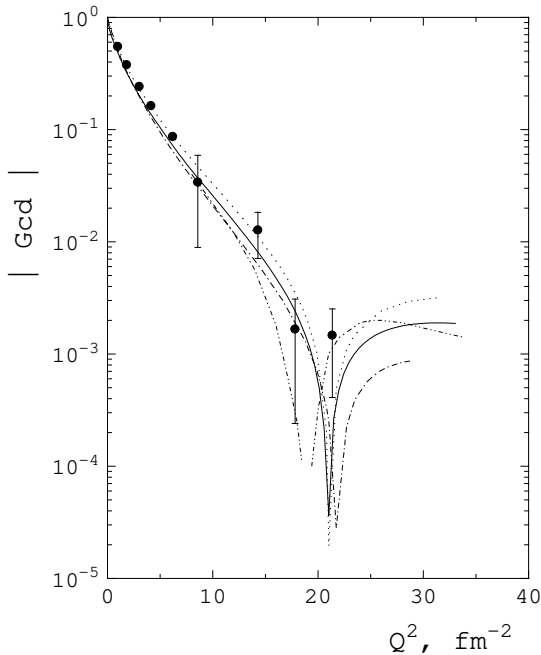}}}
\fi
\caption{Relativistic deuteron charge form factor (solid line) and its nonrelativistic
       limit (dash-dotted line). A result with nonzero values of $G_{En} = -\mu _n\tau G_{Ep}$ is
       also shown with a short-dash line. A representative result of the relativistic approach of Arnold,
       Carlson,Gross \cite{Gross} (dash-double-dotted line)  is presented for comparison.}
\label{fig2}
\end{figure}

      The result of the calculations are presented in Fig.~\ref{fig2}. Our
  brief conclusions are the following. 
  
  \noindent 1. The appearance  of zero and
  secondary maximum in $|G_{Cd}(Q^2)|$ at intermediate values of $Q^2$
   is the direct consequence of the change of sign of the experimental
  $^3S_1$-- phase shifts at intermediate energies.  It is easy to calculate
  that the model's  $\delta (E)$, which decreases
  monotonically with $E$ and is always positive ($\delta (E) > 0 $ for
  all $E$), immediately leads to monotonically decreasing with $Q^2$
  values of $|G_{Cd}(Q^2)|$  without any fine structure. 
  
 \noindent 2.  Almost up to  the
  point of zero ($Q^2 \simeq 21\; fm^{-2})$ of $|G_{Cd}(Q^2)|$  the total
  relativistic correction (TRC), {\it i.e.}, the difference between $G_{Cd}$ calculated 
  relativistically (1,2)
  and its nonrelativistic limit, is positive and appears to be not 
  small.  For example, for $Q^2 \simeq 14\; fm^{-2} $ it reaches the value
  of 25\%. 
  
\noindent  3. TRC becomes large in the region of the secondary maximum
of $|G_{Cd}(Q^2)|$, increasing the magnitude of the form factor. 

\noindent 4. The obtained results are consistent with the available data on $G_{Cd}$
from MIT-Bates \cite{bates}. Forthcoming data from Jefferson Lab E--94-018 \cite{betsy}
are extremely important to test the proposed relativistic approach in the region of higher
transferred momenta, where relativistic corrections appear to be significant.

    We would like to make the following comments to the obtained results. First,
  the dependence of $|G_{Cd}(Q^2)|$ structure on the choice of different sets of
  experimental $n-p$ phase shifts  available from the
  literature  is strong enough. Possible variation of $\delta ,
  \varepsilon , \eta $ may shift the position of zero in $|G_{Cd}(Q^2)|$
  from the indicated point $Q^2 = 21\;fm^{-2}$ to the point
 $Q^2 = 16\;fm^{-2}$ or to the point $Q^2 = 23\;fm^{-2}$. At the same time
  the secondary maximum is located in the interval $26\leq Q^2 \leq 32 \;
  fm^{-2}$, and its height may change by a factor of seven. We can see that for
  improving our understanding of $|G_{Cd}(Q^2)|$  it would be  desirable
  to obtain a more definite phase shifts analysis  of $n-p$  scattering
  in triplet channel in intermediate energy region $E_{lab} \leq 1 $ GeV.
  Secondly, let us indicate the dependence of $|G_{Cd}(Q^2)|$ on the possible choice
of nucleon EM form factors. Since the uncertainties of $G_{Ep}(Q^2)$ in the considered
range of $Q^2$ are very small, the main effect in $|G_{Cd}(Q^2)|$ may be caused only
by variation of $G_{En}(Q^2)$. It seems to be generally accepted that the
maximal deviation of $G_{En}(Q^2)$  from the zero-value  approximation
$G_{En} \equiv 0$ is given by known formula  $G_{En}(Q^2) = - \mu _n\tau
G_{Ep}(Q^2)$, where $\mu _n = -1.91$ is the neutron anomalous  magnetic
moment and $\tau = Q^2/4M^2$. The results of the calculations of
$|G_{Cd}(Q^2)|$ with this nonzero values of  $G_{En}(Q^2)$  are shown in
  Fig.2.  One can see that the effect is sizable and the contributions of
relativistic effects and nonzero $G_{En}$ have a similar behaviour.

   Finally, we show for comparison in Fig.2 the results of calculation of
$G_{Cd}$ in a relativistic approach, developed in ref.~\cite{Gross}. It may be
seen that zero of $|G_{Cd}(Q^2)|$ predicted in ref. ~\cite{Gross} is shifted to
 the lower values of $Q^2$ and
the height of the secondary maximum is approximately the same as in
our calculations. Note that in more recent calculations in the similar approach
\cite{wally} the predicted position of zero in $|G_{Cd}(Q^2)|$  remains almost
unchanged.

      Here we restricted ourselves only to the discussion of the deuteron charge
       form factor $G_{Cd}$ .
Even in this case we omitted such interesting questions as an  analytical
representation of relativistic corrections in different orders in $(v/c)^2$, the new
representation for realistic deuteron wave functions, the role of
relativistic rotation of nucleon spins and orbital momentum $l=2$ in the
deuteron, the problem of extraction, using the present approach, of
$G_{En}(Q^2)$ for ultralow values of $Q^2$ from experimental data on
elastic ed--scattering, and contributions from meson--exchange currents.
  It would also be interesting to perform a detailed
comparison of the present approach with other relativistic approaches to
the description of deuteron structure.

All these questions, as well as the calculations of the deuteron magnetic 
and quadrupole form factors will be discussed in forthcoming publications.
\bigskip

{\bf Acknowledgements}.

A.A. would like to thank F. Gross, J.W. Van Orden and I. Strakovsky for
 useful discussions. The work of A.A. was supported by
  the US Department of Energy under contract DE--AC05--84ER40150.
\end{sloppypar}

\appendix
 \section{Kinematic variables.}

      By definition $s$ is the invariant mass of $n-p$ system squared:
  \[s= (p_n + p_p)^2_\mu \:. \]
  In laboratory (LS) and center-of-mass (CMS)  systems we have
   \[ s = 4M^2+2E = 4M^2 +4p^2 \:,\]
where $E$ is the nucleon's energy in LS and $p$ is modulus of the nucleon
3--momentum in CMS.

      $Q^2$ is the magnitude of the 4--momentum transfer squared:
\[ Q^2 \equiv -q^2_\mu \equiv -t > 0 \:.\]

\section{Jost matrices $B,\tilde{B}$.}

      The formulae for pairs ($S,B$) and ($\tilde{S},\tilde{B}$) have the
most convenient form in the $p$--plane:
\[S(p)B_{+}(p) = B_{-}(p)\:,\mbox{   } -\infty < p < \infty \:,\]
where $ S\equiv S[\delta (p), \eta (p), \varepsilon (p)]$, see
eq.(\ref{four}). Let us  introduce two new matrices $\tilde{S}$ and
$\tilde{B}$:

   \[ \begin{array}{rlr}
        \tilde{B}_{\pm}(p)& = R(\mp p)B_{\pm}(p), &   \\
          R(p) &= I-{2i\alpha \over(p+i\alpha )(1+\rho ^2)}\cdot
         \left( \begin{array}{lr}
                    1&-\rho \\
                   -\rho &\rho ^2
                 \end{array}
         \right), &(\alpha ^2 = M\varepsilon) .
        \end{array}
     \]

   Now  the equation for $\tilde{B}$ has the form
\begin{equation}
  \left\{ \begin{array}{rlr}
       \tilde{S(p)}\tilde{B}_{+}(p)=&\tilde{B}_{-}(p),&-\infty<p<\infty\;,\\
       \tilde{S}(p)=&R(p)S(p)R^{-1}(-p)\equiv
       \tilde{S}[\tilde{\delta }, \tilde{ \eta },\tilde{\varepsilon }].&
       \end{array}
  \right.
  \label{b1}
  \end{equation}

   The last equation defines the reduced phase shifts $\tilde{\delta },
   \tilde{\varepsilon },\tilde{\eta}$ as functions of input experimental
   phase shifts $\delta ,\varepsilon ,\eta $.

   The solution of eq.(\ref{b1}) was found in ref.\cite{Muz} in the form of
   series
   \[
     \tilde{B}_{\pm}(p) = \tilde{B}_{\pm,0}(p)\cdot[I+\sum_{m=1}^{\infty}
     \tilde{B}_{\pm,m}(p)],
    \]
    where
  \[
   \tilde{B}_{\pm,0}(p)=
     \left(
          \begin{array}{lr}
                 \varphi _1(p)e^{\mp \tilde{ \delta }(p)}&0\\
                  0&\varphi _2(p) e^{\mp \tilde{ \eta }(p)}
           \end{array}
      \right)\:,
     \]

    \[
    \varphi _1(p)= \exp[-\frac{1}{\pi }  V.P. \int _{-\infty}^{\infty}
        {\tilde \delta (p')dp' \over p'-p}],
     \]
    \[
    \varphi _2(p)= \exp[-\frac{1}{\pi }  V.P. \int _{-\infty}^{\infty}
        {\tilde \eta (p')dp' \over p'-p}],
     \]

\begin{equation}
  \tilde B_{\pm,m}(p) =\frac{1}{2\pi i} \int_{-\infty}^{\infty} { dp'
   \over p-p' \pm i0} \cdot \sum_{n=1}^m G_n(p')\tilde B_{+,0}(p')
   [\tilde B_{+,m-n}(p')]^{1-\delta _{mn}}.
\label{b2}
\end{equation}

  In eq.(\ref{b2}) for odd $n$
  \[
    G_n(p) =i(-1)^{\frac{n-1}{2}}\cdot \frac{1}{n!}\cdot
                 [2 \tilde \varepsilon(p)]^n\cdot
                \left( \begin{array}{ll}
                       0& e^{i(\tilde \delta +\tilde \eta)} \\
                       e^{i(\tilde \delta +\tilde \eta)} &0
                        \end{array}
                 \right )
   \]
 and  for even $n$
  \[
    G_n(p) =i(-1)^{\frac{n}{2}}\cdot \frac{1}{n!}[2 \tilde \varepsilon
   (p)]^n\cdot \left( \begin{array}{ll}
                       e^{2i\tilde \delta } &0\\
                       0& e^{2i\tilde \eta}
                        \end{array}
                 \right ),
   \]
  $\delta _{mn}$ is the Kroneker delta.

\section{$g^{ll'}_c$--matrix.}
      In terms of invariant variables $s,s',t$ and the nucleon EM form factors the
matrix elements have the form:
\begin{eqnarray*}
\lefteqn{g_c^{00}(s,s',t) = }\\
&&g(s,s',t)[g_1(s,s',t)(\cos\alpha _1 \cos\alpha _2 - \frac{1}{3}
  \sin\alpha_1\sin\alpha _2)\cdot G_{EN}^s(Q^2) +\\
&&+\frac{1}{2M} g_2(s,s',t)\cdot (\frac{1}{3}\sin \alpha _1 \cos\alpha _2 -
\cos \alpha _1 \sin \alpha _2)\cdot G^s_{MN}(Q^2)] \;,
\end{eqnarray*}
\begin{eqnarray*}
\lefteqn{g_c^{02}(s,s',t) = }\\
&&g(s,s',t)\{ g_1(s,s',t)(-\sqrt{2} P_{20}\cos\alpha _1\sin\alpha _2 +
\frac{1}{\sqrt 2}P_{21} \sin\alpha _1 \cos\alpha _2 ) \cdot G^s_{EN} \\
&&-\frac{1}{2M} g_2(s,s',t) (\sqrt 2 P_{20} \sin\alpha _1 \cos\alpha _2 +
\frac{1}{\sqrt 2}P_{21} \cos \alpha _1 \sin \alpha _2 ) G^s_{MN}  \}, \\
\end{eqnarray*}
\[  g_c^{20}(s,s',t) =  g_c^{02}(s',s,t),  \]

\begin{eqnarray*}
\lefteqn{g_c^{22}(s,s',t) = }\\
&&g(s,s',t) \Bigl\{ g_1(s,s',t)
  \Bigl[(\frac{1}{3} P_{21} P'_{21} + \frac{2}{3}
   P_{20}P'_{20})\cos(\alpha _1 - \alpha _2) + \\
&& + ( \frac{1}{12} P_{22}P'_{22} + \frac{1}{3} P_{20}P'_{20})\cos\alpha _1
\cos\alpha _2 +  \\
&&+\Bigl(
          \frac{1}{12}(P_{22}P'_{21} - P_{21}P'_{22}) +
          \frac{1}{2}(P_{21}P'_{20} -P_{20}P'_{21})
    \Bigr)\sin( \alpha _1 - \alpha _2 ) - \\
&& - \frac{1}{6}(P_{22}P'_{20} + P_{20}P'_{22}) \sin\alpha _1 \sin\alpha _2
  \Bigr ]\cdot G^s_{EN} - \frac{1}{2M}g_2(s,s',t)\cdot  \\
&& \Bigl[
 \frac{1}{12}\bigl((P_{21}P'_{22} - P_{22}P'_{21}) +
 \frac{1}{2} (P_{20}P'_{21}-P_{21}P'_{20})\bigr) \cdot \\
&&\cos(\alpha _1-\alpha  _2)-( \frac{1}{12} P_{22}P'_{22} + \frac{1}{3}
P_{20}P'_{20})\cos\alpha _1\sin\alpha _2 -\\
&&-\frac{1}{6}( P_{22}P'_{20} -
 P_{20}P'_{22}) \sin\alpha _1 \cos\alpha _2 + \\
 && + ( \frac{1}{3} P_{21}P'_{21}+\frac{2}{3}P_{20}P'_{20})
  \sin ( \alpha _1 - \alpha _2)
  \Bigr] \cdot G^s_{MN}
  \Bigr\}\;,
 \end{eqnarray*}

where
\[
\begin{array}{rl}
  g(s,s',t) = & {g_1(s,s',t)\cdot(-t) \over\sqrt{(s-4M^2)(s'-4M^2)}} \cdot
   \frac{1}{[\lambda (s,s',t)]^{3/2}} \cdot \frac{1}{\sqrt{1+\tau }} ,\\
  g_1(s,s',t) = & s+s'- t,\\
  g_2(s,s',t) = & \Bigl[(-1)\bigl(M^2\lambda (s,s',t)+s s' t\bigl)
                   \Bigr]^{1/2},\\
\lambda (s,s',t) = & s^2+s'^2+t^2 -2(ss'+st+s't)  .
\end{array}
\]

      $P_{lm}$ are the Legendre polynomials, $P_{lm}\equiv P_{lm}(x)$ and
$P'_{lm}\equiv P_{lm}(x')$, where

\[
\begin{array}{rl}
x(s,s',t)=& \frac{\sqrt{s'(s'-s-t)}}{\sqrt{(s'-4M^2)\lambda (s,s',t)}},\\
 x'(s,s',t)=& -x(s',s,t).
\end{array}
\]
The angles $\alpha _1,\alpha _2$ of the relativistic rotation of nucleon
spins in deuteron are
\[
\begin{array}{ll}
\alpha _1 =& \arctan{g_2(s,s',t) \over M\bigl[(\sqrt{s} +\sqrt{s'})^2 -
              t \bigr] +\sqrt{ss'}(\sqrt{s} +\sqrt{s'} +2M)},\\
\alpha _2=& \arctan{g_2(s,s',t) (\sqrt{s} +\sqrt{s'} +2M) \over
           M(s+s'-t)(\sqrt{s} +\sqrt{s'}+2M) +\sqrt{ss'}( 4M^2-t)}.
\end{array}
 \]
     $\tau =Q^2/4M^2$ ; $G^s_{E,MN} =\frac{1}{2}(G_{E,Mp} + G_{E,Mn})$
are the nucleon isoscalar charge and magnetic form factors.


\begin{thebibliography} {10}
  \bibitem{betsy} Jefferson Lab Experiment 94-018, Measurement of the Deuteron Tensor Polarization
                 at Large Momentum Transfers in $D(e,e'd)$ Scattering, Spokepersons: S. Kox, E.J. Beise.
\bibitem{makis} Jefferson Lab Experiment 94-018, Measurement of the Electric and Magnetic Elastic
 Structure Functions of the Deuteron at Large Momentum Transfers, Spokeperson: G.G. Petratos.
 \bibitem{bates}  M. Garcon, J. Arvieux, D.H. Beck {\it et al.}, Phys. Rev.  C49 (1994) 2516.
  \bibitem{two} V.E. Troitski, S.V. Trubnikov, I.I. Belyantsev, In:
                Few Body Problem in Physics, ed. by L.D. Faddeev and
                I.I. Kopaleishvili, World Scientific, Singapore, 1985, p.480.
  \bibitem{three} V.M. Muzafarov, V.E. Troitski, S.V. Trubnikov, Sov. J. Part. Nucl.14 (5) (1983) 467.
 \bibitem{four} V.A. Romanov, S.V. Trubnikov,  R.L. Kostin, Bull. of the
                Academy of Sciences of the USSR, Phys.Ser. 53 (1989) 100.
 \bibitem{Virginia} R.A. Arndt, L.D. Roper, L.R.Workman and M.W. McNauhton,
         Phys. Rev. D45 (1992) 3995.
  \bibitem{Gross} R.G. Arnold, C.E.  Carlson, F. Gross,
       Phys. Rev. C21 (1980) 3995.
 \bibitem{wally} J.W. Van Orden, N. Devine, and F. Gross, Phys. Rev. Lett. 75 (1995) 4369.
  \bibitem{Muz} V.M. Muzafarov, Theor. Math. Phys. (USSR) 58 (1984) 184.
 \end{thebibliography}
\end{document}